\documentclass[twoside]{photon2007}
\usepackage[latin1]{inputenc}
\usepackage[dvips]{graphicx,epsfig,color}
\usepackage{wrapfig,rotating}
\usepackage{amssymb,amsmath,array}

\begin{document}
\title{
Probing Transversity GPDs in Photo and Electroproduction of Two Vector Mesons
 }
\author{
B. Pire$^{1}$ and L. Szymanowski$^{2}$
\vspace{.3cm}\\
{\it $^1$  
CPhT, \'Ecole
Polytechnique CNRS, 91128 Palaiseau, France} 
\vspace{.1cm}\\
{\it $^2$  Soltan Institute for Nuclear Studies, Warsaw, Poland and }\\
{\it University of Li{\`e}ge, Li{\`e}ge, Belgium}\\
}

\maketitle

\begin{abstract}
Electroproduction of two  mesons well separated in rapidity 
allows the
first feasible selective access to chiral-odd transversity GPDs provided one of
these mesons is a
transversely polarized vector meson $\rho_T$.
\end{abstract}

\section{Introduction}

The study of transversity \cite{transintro} is of fundamental
interest for understanding the
spin structure of nucleons. Generalized Parton Distributions (GPDs)
are the non-perturbative
objects encoding the information about the quark and gluon proton
structure in the most complete way. While the chiral even GPDs
are probed in various hard exclusive processes, the access to
transverse spin dependent
chiral-odd GPDs  has only recently been shown to be possible \cite
{transPL}.
Since in the massless quark limit,  chiral-odd functions must appear
in pairs in a non-vanishing scattering amplitude, so that chirality
flip encoded in one of them is compensated by another, a crucial
point of our proposal is to
use the transverse $\rho$ meson chiral-odd Distribution Amplitude. We
propose to use the
transverse momentum of the first meson as the large scale
needed to garanty the factorization of the GPD from a hard
subprocess. The nature of the first meson
depends on the energy range of the experiment. At large energy (e.g.
in the Compass
kinematics), diffractive production of a vector  meson is much
favored as compared
to pseudoscalar meson production, and we thus focus on the process
 \begin{equation}
 \label{transproc}
\gamma^{(*)}(q)\;  p (p_2) \to  \rho_L^0(q_\rho)\;  \rho_T^+(p_\rho)\;
n(p_{2}')
\end{equation}
while at lower energies (e.g. in the JLab kinematics), the process
\begin{equation}
\label{transproc2}
\gamma^{(*)}(q)\;  p (p_2) \to  \pi^+(q_\pi)\;  \rho_T^0(p_\rho)\;
n(p_{2}')\;,
\end{equation}
may be easier to detect. 
In both cases, one should also consider
the reference process, { \em without} the transversity GPD
\begin{equation}
\label{refproc}
\gamma^{(*)} (q)\;  p (p_2) \to  \rho_L^0(q_\rho)\;  \rho_L^+(p_\rho)\;
n(p_{2}')\;.
\end{equation}

\begin{figure}[h]
\centerline{\includegraphics[width=0.7\columnwidth]{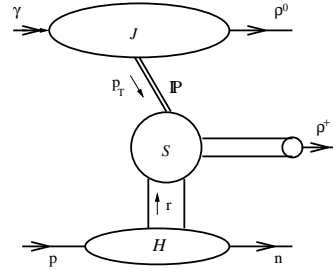}}
\caption{The process $\gamma_{L/T}^{(*)}\;  p \to  \rho_L^0\;  \rho_{L/T}^+\;
n$ in the QCD factorization approach.} 
\end{figure}
In this presentation, we restrict to the large energy case where the
off-shell exchanged object
may be seen as a Pomeron (Fig. 1) and a complete computation is
available \cite{transEPJ}. The kinematics are described by the usual
Mandelstam variables
$s=(q+p_2)^2$, $s_1=(q_\rho +p_\rho)^2$ and $s_2=(p_\rho+p_{2'})^2$
and a skewedness parameter $\xi\approx \frac{s_1+Q^2}{2s}$.
$\vec p$ is the transverse (with respect to $q$ and $p_2$) momentum
transfer in the two gluon (Pomeron) channel, Fig. 2.
We have
\begin{equation}
\label{rapgap}
s_1 \approx 2\xi s\;,~~~~s_1 >> \vec p^{\;2}\;,~~~~s_2 \approx \frac{\vec p^
{\;2}}{2\xi}(1-\xi)\;.
\nonumber
\end{equation}

\section{The scattering amplitude}

We have shown \cite{transPL} that the Born term for the process
(\ref{transproc}) is
consistently calculable within the collinear factorization method. The
amplitude is the convolution of a
hard (perturbative) part $T^q_H(x,u,z)$, two (non-perturbative) meson
DAs $\phi_{\rho^+}(u)$ and $\phi_{\rho^0}(z)$
and the (non-perturbative) GPDs of the target $H^q(x,\xi,0)$,
written as an integral over the longitudinal momentum fractions of the
quarks
\begin{eqnarray}
&&\hspace*{-0.5cm}
{\cal M} \sim \sum\limits_{q=u,d}\int\limits_0^1 dz\,\int\limits_0^1
du\,\int\limits_{-1}^1 dx\,
\nonumber \\
&&\hspace*{-0.5cm}
T^q_H(x,u,z)\,H^q(x,\xi,0) \phi_{\rho^+}(u)
\phi_{\rho^0}(z)
\end{eqnarray}
The hard part $T^q_H(x,u,z)$ is described by 6 diagrams ( Fig. 2).
and  the hard scale is the ``Pomeron'' virtuality $p^2 = p_T^2=- {\vec p}\;^2$

\begin{figure}[h]
\centerline{\includegraphics[width=0.7\columnwidth]{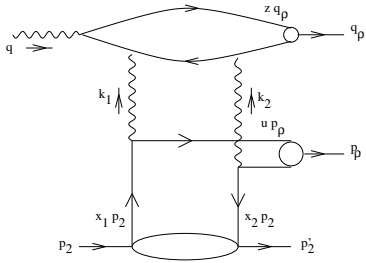}}
\caption{One of the 6 diagrams contributing to the scattering amplitude.}
\end{figure}

\vskip.1in
The  longitudinal $\rho^0_L(q_\rho)$ or  $\rho^+_L(p_\rho)$ distribution 
amplitudes (DAs)
are defined from
\begin{eqnarray}
&&\langle 0 | \bar q(-x) \gamma^\mu q(x)|\rho^0_L(q_\rho)\rangle 
\nonumber \\
&&
= q_\rho^\mu f_{\rho}^0 \int\limits_0^1du\;e^{i(1-2u)(q_\rho x)}\phi_{||}(u)
\end{eqnarray}
with
$
\phi_\|(u)=6u\bar u ~~~~~f_{\rho^0_L}=216\pm 5\,MeV
$
$
f_{\rho^+_L}=198\pm 7\,MeV 
$
while the transverse $\rho^0_T(p_\rho)$ DA is 
\begin{eqnarray}
&&\hspace*{-0.89cm}
\langle \rho_T(p_\rho,T) \mid \bar q(x) \sigma^{\mu \nu} q(-x)\mid 0
\rangle 
= i f_\rho^T 
 \\
&&\hspace*{-0.89cm}
\left(p_\rho^{\mu}\epsilon^{*\nu}_T -
p_\rho^{\nu}\epsilon^{*\mu}_T
\right)
\int\limits_0^1 du e^{-i(2u-1)(p_\rho x)}\;\phi_\perp(u)
\nonumber
\end{eqnarray}
with$
\phi_\perp(u)=6u\bar u ~~~~~~~~~f_{\rho^+_T}=160\pm 10\,MeV
$

The transversity GPD is a non diagonal matrix element of the nonlocal operator
\[
\hat O_T =\bar q(-\frac{z}{2})\, i \sigma^{+\,i}\, 
q(\frac{z}{2})
\]
\begin{eqnarray}
&&\hspace*{-0.99cm}
\int \frac {dz^-}{4\pi}
e^{i x P^+ z^-}
\langle N(p_{2'},n)| \hat O_T| N(p_2,n) \rangle
 \\
&&\hspace*{-0.99cm}
=\frac{1}{2P^+}\bar u(p_{2'},n) i \sigma^{+ \,i} u(p_2,n) H_{T}^q(x,\xi,t) + ...
\nonumber
\end{eqnarray}

Not much is known about such a non perturbative object, and different plausible models 
will yield quite different counting rates. We have used two models: we generalized the 
meson pole model by Gamberg et al \cite{gamberg} to the kinematics with GPDs
and a model by Scopetta \cite{scopetta} based on the bag model. 
The resulting GPDs  are very
different as seen on  Fig. 3.

\begin{figure}[h]
\centerline{\includegraphics[width=1.0\columnwidth]{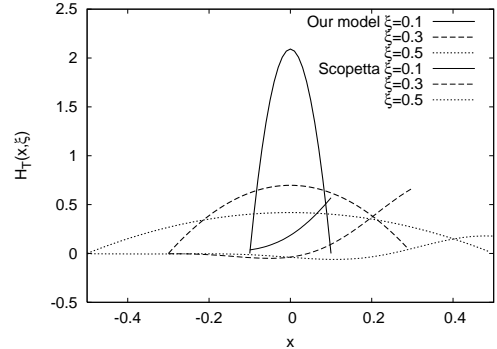}}
\caption{The transversity GPD $H_T$ in the ERBL region in the meson pole 
model and in the model of Ref. \cite{scopetta}.}
\end{figure}

The reference process with $\rho^+_L$  involves usual (but still largely unknown)
non-polarized nucleon GPDs, constructed from the operator
$
\hat O= \bar q(-\frac{z}{2})\,  \gamma^{+}\, 
q(\frac{z}{2})
$

\section{Results and conclusions}
 Fig. 4 and in Fig. 5 show the photoproduction cross sections
for the reference process (\ref{refproc}) and 
for process  (\ref{transproc}) with the
 transversity GPD described by the meson pole model.
\begin{figure}[h]
\centerline{\includegraphics[width=1.0\columnwidth]{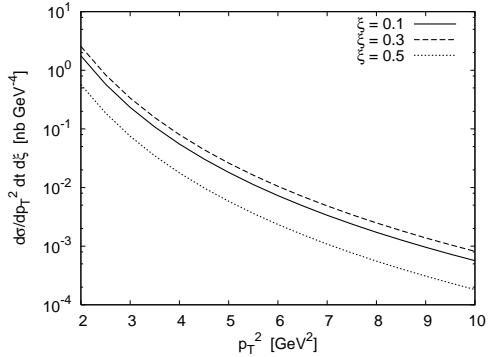}}
\caption{The cross-section for the reference process of 
photoproduction of $\rho^0_L$ and $\rho^+_L$. }
\end{figure}

\begin{figure}[h]
\centerline{\includegraphics[width=1.0\columnwidth]{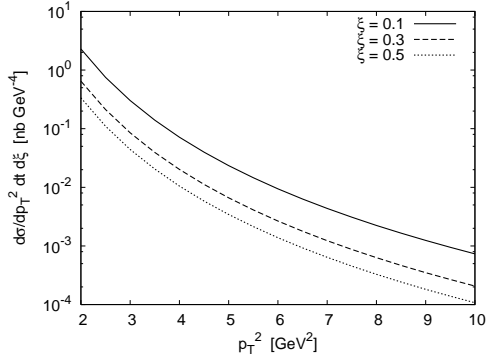}}
\caption{
The cross-section for the  
photoproduction of $\rho^0_L$ and $\rho^+_T$ with the transversity GPD 
modeled by the $b_1$ meson exchange.
}
\end{figure}

We see that the magnitudes of the cross sections are quite sizable in both cases,
and we can thus infer that an experiment like 
COMPASS can measure these processes and thus access for the first time the transversity GPD.

Let us note that by replacing in the above process $\rho_T^0$ by a real photon $\gamma$ 
and using the chiral-odd photon DA one may probe 
the magnetic susceptibility $\chi$ of the QCD vacuum. Indeed the photon DA is
 defined by a similar correlator
as the transversity GPD \cite{magsusc}
\begin{eqnarray}
&&
\hspace*{-1.1cm}
\langle 0 |\bar q(0) \sigma^{\alpha\beta}\,q(x)|\gamma^{(\lambda)}(p)\rangle = 
ie_q\,\chi \,\langle \bar q q \rangle
 \\ 
&&
\hspace*{-1.1cm}
\left( \epsilon^{(\lambda)}_\alpha p_{\beta} - \epsilon^{(\lambda)}_\beta p_{\alpha} \right) \int\limits^1_0\;du\,e^{-iu(px)}\phi_\gamma(u,\mu)
\nonumber
\end{eqnarray}
where
$
\chi \,\langle \bar q q \rangle \approx 40 - 70 \mbox{MeV}
$

\section*{Acknowledgements}
We acknowledge common research  with R. Enberg, D.Yu. Ivanov and O.V. Teryaev.
This work was supported in part by the Polish Grant 1 P03B 028 28.
L.Sz. is a visiting Fellow of the FNRS (Belgium).


\begin{footnotesize}



%

\end{footnotesize}


\end{document}